\documentclass[12pt]{article}
\usepackage[utf8]{inputenc}
\usepackage{graphicx}
\usepackage{geometry}
\usepackage{array}
\usepackage[svgnames,table]{xcolor}
\newcommand*{\arraycolor}[1]{\protect\leavevmode\color{#1}}
\newcolumntype{A}{>{\columncolor{blue!50!white}}c}
\newcolumntype{B}{>{\columncolor{LightGoldenrod}}c}
\newcolumntype{C}{>{\columncolor{FireBrick!50}}c}
\newcolumntype{D}{>{\columncolor{White}}l}
\newcolumntype{E}{>{\columncolor{Gray!42}}l}

\usepackage{xcolor}
\definecolor{Green}{rgb}{0,0.5,0}
\definecolor{Blue}{rgb}{0,0,1}
\definecolor{Red}{rgb}{1,0,0}

\begin{document}
\thispagestyle{empty}
\Large
\bf 

 Integrating Undergraduate Research and Faculty Development in a Legacy Astronomy Research Project
\rm

\vspace{0.1in}

Astro2020 APC White Paper 
\rm

\noindent \large Thematic Area: \bf State of the Profession\\
\rm
\vspace{0.1in}
\normalsize

\textbf{Lead Author}:\\
Rebecca A. Koopmann, Union C., koopmanr@union.edu

\vspace{0.1in}

\textbf{Co-Authors for the Undergraduate ALFALFA Team}:\\
Thomas J. Balonek, Colgate University, tbalonek@colgate.edu\\
John M. Cannon, Macalester College, jcannon@macalester.edu\\
David Craig, West Texas A \& M University, dcraig@wtamu.edu\\
Adriana Durbala, University of Wisconsin-Stevens Point, adurbala@uwsp.edu\\
Rose Finn, Siena College, rfinn@siena.edu \\
Gregory Hallenbeck, Washington \& Jefferson College, ghallenbeck@washjeff.edu\\
Martha Haynes, Cornell University, haynes@astro.cornell.edu\\
Mayra Lebr\'on, University of Puerto Rico - R\'\i o Piedras, mayra.lebron3@upr.edu\\
Lukas Leisman, Valparaiso University, luke.leisman@valpo.edu\\
Jeffrey Miller, St. Lawrence University, 
Mary Crone Odekon, Skidmore College, mcrone@skidmore.edu\\
Aileen O'Donoghue, St. Lawrence University, aodonoghue@stlawu.edu\\
Joseph Ribaudo, Providence College, Utica College, jribaudo@providence.edu\\
Jessica Rosenberg, George Mason University, jrosenb4@gmu.edu\\
Parker Troischt, Harwick College, TroischtP@hartwick.edu\\
Aparna Venkatesan, University of San Francisco, avenkatesan@usfca.edu\\

\normalsize

\newpage
\clearpage
\setcounter{page}{1}

\bf \large Executive Summary\rm \normalsize

This White Paper addresses the importance of 
engaging faculty and students at primarily undergraduate institutions (PUI) in frontier astronomy research. The current astronomy research landscape focuses on high intensity engagement, leaving unclear the roles of many at PUIs who have the training, desire, and skills to contribute, as well as access to a diverse pool of citizens and potential scientists (see also White Paper by Ribaudo et al.). 
The challenge for faculty at the PUIs can be in finding opportunities to contribute their research skills and interests that match the level of intensity appropriate for their diverse circumstances.

We describe our model, the NSF-sponsored Undergraduate ALFALFA Team (UAT), founded to promote long-term collaborative research opportunities for faculty and students from a wide range of public and private PUIs
within the context of the extragalactic ALFALFA HI blind legacy survey project. 
Over twelve project years of partnering with Arecibo and Green Bank Observatories, the UAT has demonstrated 
success in peer mentoring and training, enhancing faculty and student research in small physics and astronomy programs, student pathways to STEM careers, and inclusion of more women and minorities in leadership roles. 
These are known factors that improve long-term diversity and productive outcomes in STEM fields
(Phillips 2014; White Paper by Norman et al.).

This modest investment from the NSF has had a strong, demonstrable impact on the health of a legacy astronomy project, science education, and equity/inclusion in astronomy, with
successful outcomes for 
373 UAT students (39\% women; $\sim$30\% members of underrepresented groups) and 34 faculty (44\% women). 
While the federal agencies currently offer programs such as REUs to promote early-career involvement in research, the long-term outcomes for diversity/inclusion and retention beyond individual summers or grant years are not clear, especially for underrepresented populations (e.g., Slater 2010). 
The UAT model, on the other hand, offers a productive avenue for faculty at PUIs, 
where Ph.D.s in astronomy are increasingly employed (see White Paper by Ribaudo et al.), to sustain successful research programs and contribute to national collaborations through collective effort.

\vspace{0.1in}

\bf \large Recommendations:\rm\normalsize
\vspace{-0.1in}
\begin{itemize}

\item That large scientific projects and research collaborations recognize and better utilize the high numbers of heavy-teaching-load faculty who are astronomy research-trained as a valuable resource to the astronomy community - both for the insights they can bring to research and their access to a diverse pool of potential early-career scientists.

\item 
That the astronomy community identify and create funding resources and mechanisms to support programs such as the UAT, either as add-on funding to legacy grant support or as stand-alone funding sources. This could include funding UAT-like components in large-scale long-term projects such as the LSST and TMT.

\item That the NSF and other agencies increase investment in long-term scientific projects that: 1) intentionally focus on (and demonstrate sustained success in) retaining students at early career stages through direct involvement in frontier scientific projects and through peer mentoring, and 2) promote women and underrepresented minorities in leadership positions.

\end{itemize}

\newpage

\section{Key Issue and Overview of Impact on the Field}

The migration of astronomy research toward large collaborations with the work extending over long time periods has made it increasingly challenging for undergraduates and faculty at Primarily Undergraduate Institutions (PUIs)
to contribute to basic astronomy research. Undergraduate projects are often limited to a summer or an academic term, and undergraduates often begin their projects with minimum background coursework in astronomy. Faculty at institutions with heavy teaching loads often have limited time to participate (mainly summers and sabbaticals), but they have a strong desire to contribute and their expertise and training are resources that are valuable to a research team.

Effective involvement of undergraduates in research has been shown to inspire students to continue in STEM fields (Russell et al. 2007; Eagan et al. 2013), which is ever more important as we see a declining number of domestic students continuing in science (e.g., National Academies 2007). Numerous studies have found that
the undergraduate research experience is a high-impact pedagogical practice (Sadler et al. 2010) that should be available to as broad a range of students as possible (e.g., Kuh 2008; Rowlett, Blockus, and Larson 2012). 

The importance of having professionally engaged teaching faculty who belong to “communities of practice” is also well documented (e.g., Wenger 1998). Faculty at PUIs often teach  introductory courses which may be the only place where
 non-STEM students encounter research in the physical 
 sciences. 
 Faculty scholarship informs these courses both in content and in the discussion of research from the perspective of a faculty member who is a participant rather than simply a reporter.
 Interactions between faculty and students in a research setting differ from those in the classroom and laboratory because they are more collaborative than instructive and model a more authentic practice of science than the ones that most undergraduate students experience.  These broaden students' experience of science as they interact with faculty and other students in actual scientific research during the academic year and in summer. This enhances and extends students' experience of research beyond what is generally possible in the NSF Research Experience for Undergraduates (REU) programs.

Our field must be creative in efforts to train the next generation of astronomers, while maintaining the contributions of our undergraduate-focused faculty. The model we describe here can be adapted to many projects and implemented at a reasonable cost.

\section{The UAT Model and Outcomes}

For the past 12 years, we, the Undergraduate ALFALFA Team (UAT), have developed a model to facilitate participation of 
undergraduates and their faculty in a radio astronomy legacy project.  The ALFALFA survey (Giovanelli et al. 2005; Haynes et al. 2018), originated and coordinated by Riccardo Giovanelli and Martha Haynes at Cornell University, mapped 7074 square degrees of the high galactic latitude sky visible from Arecibo (with many of the observing runs performed by UAT collaborators). ALFALFA provides an extragalactic HI line spectral database of 31,500 extragalactic HI sources covering the redshift range between -1600 km/s and +18,000 km/s (Haynes et al. 2018). The combination of sensitivity, bandwidth and sky coverage makes ALFALFA the first wide-area HI survey to sample a cosmologically-significant volume, resulting in 108 team publications to date.

From the outset of ALFALFA, the PIs sought to open the project to undergraduate participation. The UAT was conceived in 2004 by a small group of faculty members from several New York and Pennsylvania undergraduate liberal arts institutions, plus the University of Puerto Rico. One-day workshops were hosted at Union College in Schenectady, New York, which introduced attendees to ALFALFA and the possibility of working on related research while maintaining positions at primarily small, somewhat isolated undergraduate institutions.
The success of these workshops inspired an NSF proposal to form an expanded UAT program with a workshop at Arecibo. Now in our third term of NSF-funding, the Team has grown to include 23 public and private mostly PUI from across the U.S., including three minority serving institutions, as listed in Table 1.  Initially focused on the ALFALFA survey itself, Team members are currently working on several related followup projects.

\begin{table}[h!]
\begin{center}
\scriptsize
\arrayrulewidth=1pt
\arrayrulecolor{black}
\renewcommand{\arraystretch}{1.5}
\begin{tabular}{|D|D|D|}
\rowcolor{Black!50!White}
 \arraycolor{White}\bf Institution &  
 \arraycolor{White} \textbf{Location} &  
 \arraycolor{White} \textbf{Team Lead} \\
\rowcolor{Black!2!White}
\bf Colgate University & \bf Hamilton, NY & \bf Thomas Balonek \\
\rowcolor{Black!10!White}
\bf Cornell University & \bf Ithaca, NY & \bf Martha Haynes \\
\rowcolor{Black!2!White}
\bf George Mason University & \bf Fairfax, VA & \bf Jessica Rosenberg \\
\rowcolor{Black!10!White}
\bf Hartwick College & \bf Oneonta, NY & \bf Parker Troischt \\
\rowcolor{Black!2!White}
\bf Lafayette College & \bf Easton, PA & \bf G. Lyle Hoffman \\
\rowcolor{Black!10!White}
\bf Lynchburg College & \bf Lynchburg, VA & \bf Crystal Moorman\\
\rowcolor{Black!2!White}
\bf Macalester College & \bf St. Paul, MN & \bf John Cannon  \\
\rowcolor{Black!10!White}
\bf Metropolitan State \bf University of Denver& \bf Denver, CO & \bf Grant Denn\\
\rowcolor{Black!2!White}
\bf St. Lawrence University & \bf Canton, NY & \bf Aileen O'Donoghue \\
\rowcolor{Black!10!White}
\bf San Francisco State University$^{\dagger}$ & \bf San Francisco, CA & \bf Kim Coble \\
\rowcolor{Black!2!White}
\bf Siena College & \bf Loudonville, NY & \bf Rose Finn  \\
\rowcolor{Black!10!White}
\bf Skidmore College & \bf Saratoga Springs, NY & \bf Mary Crone Odekon \\
\rowcolor{Black!2!White}
\bf Union College & \bf Schenectady, NY & \bf Rebecca Koopmann  \\
\rowcolor{Black!10!White}
\bf University of Puerto Rico- R\'\i o Piedras$^{\dagger}$& \bf San Juan, PR & \bf Mayra Lebron Santos \\
\rowcolor{Black!2!White}
\bf University of San Francisco$^{\dagger}$ & \bf San Francisco, CA & \bf Aparna Venkatesan \\
\rowcolor{Black!10!White}
\bf University of Wisconsin - Madison & \bf Madison, WI & \bf Eric Wilcots \\
\rowcolor{Black!2!White}
\bf University of Wisconsin - Platteville & \bf Platteville, WI & \bf Kathryn Rabidoux \\
\rowcolor{Black!10!White}
\bf University of Wisconsin - Stevens Point & \bf Stevens Point, WI & \bf Adriana Durbala \\
\rowcolor{Black!2!White}
\bf Utica College& \bf Utica, NY & \bf Joseph Ribaudo \\
\rowcolor{Black!10!White}
\bf Valparaiso University & \bf Valparaiso, IN & \bf Lukas Leisman \\
\rowcolor{Black!2!White}
\bf Washington \& Jefferson College & \bf Washington, PA & \bf Gregory Hallenbeck\\
\rowcolor{Black!10!White}
\bf West Texas A\&M University & \bf Canyon, TX & \bf David Craig \\
\rowcolor{Black!2!White}
\bf West Virginia University & \bf Morgantown, WV & \bf D. J. Pisano  \\
\hline
\end{tabular}
\caption{UAT Institutions and Lead Faculty ($^{\dagger}$Minority Serving Institution as listed by the Rutgers Center for Minority Serving Institutions)}
\label{institutions}
\end{center}
\end{table}

We have identified several priorities to address the challenges of connecting to isolated researchers and their students in PUI environments: (1) communicating a basic understanding of radio astronomy and extragalactic astrophysics to students with little or no class preparation, (2) providing resources for faculty and students to obtain computers and travel funds to support research, (3) identifying undergraduate-appropriate research projects, and (4) demonstrating modern collaborative science practices.

We measure the effectiveness of the program partly through the paths taken by each student in the program after she or he graduates. Additionally, in the last seven years, our external assessor has gathered data in the form of anonymous surveys of students who attended the annual workshops, students who observed onsite at Arecibo or the Kitt Peak National Observatory (KPNO), and student alumni of the UAT program. The student surveys were adapted from the Survey on Undergraduate Research (SURE; Lopatto 2008), which identifies and measures a number of benefits expected from students’ participation in research.

The cornerstone of the UAT is the annual workshop which has been held eight times at Arecibo Observatory and four times at Green Bank Observatory, most recently in June 2019. 
The workshop provides an overview of the UAT scientific work, as well as training in research methods, as it nurtures peer and mentor collaboration. Teams of undergraduates and faculty members engage in active group projects, observe onsite, carry out hands-on data analysis, and present results. Few undergraduates have opportunities to visit a national observatory, particularly a radio astronomy observatory. The active nature of the sessions allows the students to gain an understanding of radio astronomy observations that they would not have achieved working in isolation at their home institutions. The immersive workshop experience encourages students to pursue UAT research projects on their home campuses, and, in fact, 80\% of participants have gone on to complete at least one independent research project for course credit or using UAT research funds or institutional resources. 

On-site observing sessions provide additional opportunities for students to visit national observatories (Arecibo, KPNO). 
Since the completion of ALFALFA observations in 2012, UAT faculty have collaborated with ALFALFA principal investigators to write their own proposals (Arecibo, Green Bank, Kitt Peak, VLA) to perform targeted follow-up observations, both at the workshops and during on-site or remote observing runs each semester.
Remote observing runs at Arecibo and Green Bank were made possible by  
the on-site observing training of faculty, and this opportunity has extended the
observing experience to laboratories and classrooms at multiple institutions and allowed more students and non-UAT affiliated faculty on campus to participate.

During summers and throughout the academic year, faculty mentors supervise undergraduates to work on intellectually engaging research projects covering a range of topics associated with Team projects, relying on the expertise and interest of each faculty member. Collaborative team projects, including the Arecibo Pisces Perseus Supercluster Survey (APPSS; O'Donoghue et al. 2018) and a survey of star formation and gas properties of Groups and Clusters (Odekon et al. 2016, 2018), link students and faculty at team schools to work together on basic research problems (mass of a supercluster, evolution of galaxies in groups).  
Funds to purchase computers available in our first two rounds of funding facilitated these projects, particularly at institutions unable to provide research computers for their faculty and students. The Team also provides funding for students and faculty 
to present their results at national meetings.

So far, 373 students have participated in the program, with women making up about 40\% of participants (see Table 2). In the last seven years, 31\% of participants in workshop and on-site observations self-reported themselves to be
either Hispanic/Latino (25\%) students or African-American (6\%) students. 
Participating students are chosen in a variety of ways at their home institutions. Most have taken physics, but there are no specific academic requirements for
participation; the open, usually non-competitive opportunities for students to join UAT may enhance the involvement of underrepresented groups. A few participants have even been advanced high school students enrolled in college courses. Many students stay involved over multiple years and work on more than one project, often serving as mentors to their younger colleagues. UAT faculty members have supervised 233 summer research projects and 218 academic-year projects. A total of 116 students have presented or co-authored posters at the AAS, with many others presenting at other meetings, including the American Physical Society, the National Society of Black Physicists, the Astronomical Society of New York, 
and Conferences for Undergraduate Women in Physics. Almost all student researchers have presented at on-campus venues. Some have been coauthors on journal papers (e.g. Ball et al. 2018; McNichols et al. 2016; Odekon et al. 2016, 2018).

\begin{table}[h!]
\begin{center}
\arrayrulewidth=1pt
\arrayrulecolor{black}
\renewcommand{\arraystretch}{1.5}
\rowcolors[\hline]{3}{.!50!White}{}
\begin{tabular}{|C|C|C|}
 \rowcolor{Maroon!75!White}
 \arraycolor{White}\textbf{UAT Activity} &  
 \arraycolor{White} \textbf{\# Participants} &  
 \arraycolor{White} \textbf{\% Women} \\
\rowcolor{Maroon!5!White}
\color{Maroon!50!Black} \bf Arecibo/Green Bank Workshop & \bf 214 & \bf 39 \\
\rowcolor{White}
\bf Academic Year Projects & \bf 218 & \bf 39 \\
\rowcolor{White}
\bf Summer Projects & \bf 233 & \bf 35\\
\rowcolor{Blue!5!White}
\color{Blue!50!Black} \bf Observing Run at Arecibo/Kitt Peak & \bf 109 & \bf 42\\
\rowcolor{Green!5!White}
\bf AAS Presentations (First Author) & \bf 93 & \bf 40\\
\rowcolor{Orange!25!White}
\bf Total Individuals & \bf 373 & \color{Maroon} \bf 39\\
\rowcolor{Black!5!White}
\bf Graduate School, STEM & \bf 149 & \bf 37 \\
\rowcolor{Black!5!White}
\bf Graduate School, PHY/AST & \bf 104 & \bf 38 \\
\rowcolor{Black!5!White}
\bf STEM - Employed & \bf 77 & \bf 38\\
\end{tabular}
\caption{UAT Activities, Participation, and Outcomes}
\label{tab:tab1}
\end{center}
\end{table}

Direct interaction with faculty and peer mentoring are key intrinsic parts of this program and effectively reach the next generation of scientists. Through these interactions, students are made aware of career paths they may not have previously considered. UAT alumni report that their experience provided the confidence and connections they needed to make decisions concerning their future lives, with 75\% responding that the experience increased their interest in research very much or quite a bit. Many respondents noted that UAT research was their first exposure to scientific research, and it was an intensive experience that has stuck with them throughout their careers. 80\% stated that their UAT experiences had helped them to know themselves and their interests very much or quite a bit, with several respondents noting that it had been particularly useful that they came to these realizations about their interests earlier than other students and were able to make more informed choices of future research and graduate study.
Many students have commented about the importance of the UAT experience in their decision to pursue advanced study. As of Fall 2019, 149 (37\% women) UAT alumni are or have been enrolled in graduate programs in astronomy, physics, and other science and math disciplines (including archaeology, aeronautics, architecture, atmospheric science, biology, geobiology, geoscience, nanoscience, material science, applied math, systems modeling, medical physics, and science writing). A few 
have since completed their Ph.D.s and are pursuing careers as astronomers.
Several of our institutions have had students who worked in our program go on to graduate school, after decades of no students attending graduate school. 
In total 91\% of our 248 alumni with known status are
pursuing a career in a
STEM-focused field, with those not attending graduate school pursuing
careers such as teaching, engineering/industry, software development, and medical fields. (See Table 2.)

The UAT has had a substantial impact on the professional development of UAT faculty, who might otherwise struggle to continue contributing to research due to being isolated at their home institutions. To date, 34 faculty have participated, 44\% women. The collaborative nature of the UAT provides a support structure for faculty at small colleges that maximizes the time that they can devote to the project and develops
their research and education  skills. Most of the faculty members of the UAT were new to the collaboration and many had expertise in fields outside extragalactic radio astronomy. Over the 12 years of funding, they have benefited from the opportunity to learn about and participate in the project, bringing their own sets of skills, experience, and expertise to the collaboration.  
Faculty also have used their involvement in the UAT and ALFALFA to enhance their curricular and public outreach offerings, raising awareness of astronomy on their local campuses and in their local communities. Faculty invite students and members of the public to participate in remote observing sessions, extending access to a national observatory to a wider audience on campus and the public. 
Educational materials designed to introduce students to the project are publicly available for wider distribution on the UAT and ALFALFA web pages. 
Ten faculty members have been promoted to Associate Professor, and five to Professor. A number of our faculty have received awards for research or teaching, and had significant grant awards beyond their UAT work:
two faculty have received NSF Career Grants, another received course release through an NSF ADVANCE grant, another received an NSF-RUI grant, and a fifth was named a Cottrell Scholar by Research Corporation.

Finally the UAT has impacted graduate students at several of our institutions, including Cornell, U.W. Madison, and George Mason U.
Graduate students
have been an integral part of the UAT since its inception. They contribute by giving lectures, leading small groups of students in activities, discussing their experiences in graduate school with the undergraduates, and helping to train faculty and undergraduates to observe. 
Those who have completed their Ph.D.s have all gone on to either postdoctoral or professor positions. In fact, the faculty leads at
two recently-added UAT institutions (Valparaiso, Washington \& Jefferson)  were Cornell graduate students who had participated in our program. The opportunity for graduate students to work with undergraduates and faculty at a variety of schools has proven to be invaluable in their careers and has provided accessible role models to the undergraduate participants.

\section{Applying the Model to Other Projects}

While the UAT was developed for a specific project, there are many elements that would transfer to any research project and funding level (see also Troischt et al. 2016). 
Adapting the model to other research projects provides a natural way to include broader impacts (required by most granting agencies).

The success of our program is closely tied to dedicated leadership by the ALFALFA PIs, especially Martha Haynes at Cornell University, and a designated facilitator (Rebecca Koopmann at Union College, PI of the grants supporting the program). These leaders invest a significant portion of their professional energy to run the program. The background requirements for collaborating faculty members can be very flexible. We identified UAT collaborators through existing personal connections, initially through geographical proximity to Cornell University and participation in the Astronomical Society of New York, a state-wide society for professional astronomers.  
Through the years of the UAT, however, it has expanded to include institutions across the country (Table 1) through the personal connections of faculty within the UAT and UAT graduate students becoming faculty.  The UAT now has member institutions in California, Colorado, Indiana, Minnesota, New York, Pennsylvania, Texas, Virginia, Wisconsin, and Puerto Rico, including three minority-serving institutions.
While geographical proximity can be an advantage, the advent of teleconferencing tools and cloud-shared sites enables close, regular collaboration between distant sites even when funding is limited.
The most important qualification of participating faculty is not their scientific specialty, but their willingness to learn and contribute. Our faculty members have a range of expertise; while some were trained in extragalactic radio astronomy, others are experts in fields including optical astronomy, x-ray astronomy, and string theory. We also have created a flexible
structure, where faculty, many with heavy responsibilities in teaching and administration at their home institutions, can participate in the ways that work best for them. Faculty can choose if and when to take on summer research students, help with observing runs, and attend workshops. The level of knowledge and types of participation for students are also flexible. Some participate peripherally by helping out at observing sessions or writing some computer programs, while others make significant contributions through summer and/or academic year research projects, many of which serve as “capstone” experiences required for their degrees and lead to refereed publications. We believe this flexibility has contributed to the longevity, expansion, and sustainability of the program.

Our experience shows the importance of regular communication and events to the coherence of the program. Besides the annual workshop at Arecibo or Green Bank, we have monthly teleconferences for all UAT faculty members,  research teleconferences for faculty and students, and groups that meet regularly for observing and planning to attend conferences. To ease communication among the institutions, we have both a public web site (http://egg.astro.cornell.edu/alfalfa/ugradteam/ugradteam.php) and a private Google site for background information, computer programs, sets of instructions for observing routines, and data-analysis procedures. Having very well-documented and easily accessible information allows participants to build efficiently on previous work.

Last, we emphasize one of the key strengths of our model relative to some of the other successful programs in astronomy offered by federal agencies, including the more traditional REU programs. These other programs do promote early-career involvement in research and exposure to hands-on work in leading scientific projects (Sadler et al. 2010). However, the long-term outcomes for individual students beyond individual summers or a 2-3 year grant period are not clear. For example, Slater (2010) conducted a longitudinal, two-stage study of 51 women who participated in an astronomy REU and 
found that the experience had relatively little impact on their
pursuit of astronomy, largely because successful REU applicants already knew they wanted to pursue a career in STEM. Their decision to pursue an REU and STEM career was instead far more affected by previous long-term relationships with mentors, a key element in our approach. 
In our model, rather than a 10-week experience away from their campus, students can become involved at early career stages 
in a long-term scientific project through their local faculty member.
This naturally fosters long-term peer and advisor mentoring, while providing access to a national network of faculty and students at different institutions and astronomy career stages and networking with significant numbers of women and underrepresented minorities in leadership roles. We believe that these factors, especially the long-term mentoring and sustained networking, are crucial in building confidence and community, and ultimately retention, for underrepresented populations in astronomy.

\section{Cost Estimates and Importance of Funding}

Our modest NSF funding has made possible our transformative workshops and observing experiences at national observatories, as well as summer research and conference presentations for a large number of students. We obtained this funding via unsolicited proposals to
Special Programs in Astronomy Education. While past grants provided
$\sim$ \$450k range for three years, tightened AST funding necessitated reducing the budget of our most recent grant to $\sim$\$250k for three years. Our current grant meets expenses for our
workshop and travel to conferences and observing, but has limited funds
for summer and academic year student stipends, which are especially critical
for students from disadvantaged backgrounds, and no funds for computer purchases. 
While aspects of our model can be applied with limited funding to achieve some of the goals (e.g., work can be done remotely via teleconferencing, especially for existing scientific projects, and small local workshops are possible with little funding), sufficient (and still modest) funding is necessary to carry out our work meaningfully, especially at small
colleges and state universities with tight budgets.

While this project has been successful in obtaining NSF support, \bf one of the major challenges to its continuation and expansion into other research areas is a lack of a structure for support. \rm Special projects support is limited and projects like this do not naturally fit under the standard Astronomy and Astrophysics Research grants nor do they fit under the Improving Undergraduate STEM Education support which focuses on studying the innovations themselves. New structures are needed if innovative models like the UAT are to survive and thrive.

\section{Conclusion}

The collaborative UAT model has been extremely successful at allowing faculty and students from a wide range of colleges, especially those with small astronomy programs, to develop impactful scholarly collaborations. Providing research experiences to faculty and students at 23 distinct institutions across the country has allowed us to reach the diversity of talent that is needed for our future leadership in science. The close faculty and peer relationships and networking across the country, including faculty at Research 1 institutions, provides a community for undergraduates that can influence their decision to pursue a STEM-oriented field.

Our model is adaptable to many large scientific projects and can be
supported by relatively modest funding.
We recommend that granting agencies identify  funding  resources  to  support the model, either as  an  add-on to legacy  grant
support or as a stand-alone funding source.  This could include encouragement of UAT-like components in large scale projects currently being developed, such as the LSST and TMT. By doing this, we will recognize the high numbers of astronomy research-trained heavy-teaching-load faculty at PUIs
as an under-utilized resource of the astronomy community (see also White Paper by Ribaudo et al.). These members of our community have the skills and the strong desire to contribute meaningfully to their field, as well as the ability to encourage and interact closely with many talented and motivated undergraduate students from all backgrounds. We will also reach a talented and diverse pool of future citizens and potential future scientists.

\vspace{0.2in}

\newpage
\bf{References}
\rm

Ball, C., Cannon, J.M., Leisman, L., Adams, E.A.K., Haynes, M.P., Józsa, G.I.G., McQuinn, K.B.W., Salzer, J.J., Brunker, S., Giovanelli, R., Hallenbeck, G., Janesh, W., Janowiecki, S., Jones, M.G. \& Rhode, K.L. 2018 AJ, 155, 65

Eagan, M. K., Jr, Hurtado, S., Chang, M. J., Garcia, G. A., Herrera, F. A., \& Garibay, J. C. 2013, Am Educ Res J., 50, 683 

Giovanelli, R., Haynes, M.P., Kent, B.R., Perillat, P., Saintonge, A., Brosch, N., Catinella, B., Hoffman, G.L., Stierwalt, S., Spekkens, K., Lerner, M., Masters, K.L., Momjian, E., Rosenberg, J., Springob, C.M. plus 25 others, 2005a, AJ, 130, 2598.

Haynes, M.P., Giovanelli, R., Kent, B.R., Adams, E.A.K., Balonek, T.J., Craig, D.W., Fertig, D., Finn, R., Giovanardi, C., Hallenbeck, G., Hess, K.M., Hoffman, G.L., Huang, S., Jones, M.G., Koopmann, R.A., Kornreich, D.A., Leisman, L., Miller, J., Moorman, C., O'Connor, J., O'Donoghue, A., Papastergis, E., Troischt, P. Stark, D. \& Xiao, L. 2018 ApJ, 861, 49

Kuh, George D. 2008, \it High-Impact Educational Practices: What They Are, Who Has Access to Them, and Why They Matter, \rm  Washington, DC: Association of American Colleges and Universities.

Lopatto, D. 2008, in Creating Effective Undergraduate Research Programs in Science, R. Taraban \& R.L. Blanton (Eds.), (NY: Teacher’s College Press), 112

McNichols, A.T., Teich, Y.G., Nims, E., Cannon, J.M., Adams, E.A.K., Berstein-Cooper, E.Z., Giovanelli, R., Haynes, M.P., Józsa, G., McQuinn, K.B.W., Salzer, J.J., Skillman, E.D., Warren, S.R., Dolphin, A., Elson, E.C., Haurberg, N., Ott, J., Saintonge, A., Cave, I., Hagen, C., Huang, S., Janowiecki, S., Marshall, M.V., Moody, S., Thomann, C.M., \& Van Sistine, A. 2016, ApJ, 832, 89

National Academy of Sciences, National Academy of Engineering, and Institute of Medicine 2007. \it Rising Above the Gathering Storm: Energizing and Employing America for a Brighter Economic Future, \rm (Washington, DC: The National Academies Press)

Norman, D. et al. 2019, "Tying Research Funding to Progress on Inclusion," State of the Profession White Paper, Astro 2020 Decadal Survey

Odekon, M. C., Koopmann, R., Haynes, M.P., Finn, R.A., McGowan, C., Micula, A., Reed, L., Giovanelli, R., \& Hallenbeck, G. 2016 ApJ, 824, 110 

Odekon, M. C. , Hallenbeck, G., Haynes, M.P., Koopmann, R.A., Phi, A. \& Wolfe, P.-F., 2018, ApJ, 852, 142

O'Donoghue, A.A., Haynes, M.P., Koopmann, R.A., Jones, M.G., Giovanelli, R., Balonek, T.J., Craig, D.W., Hallenbeck, G.L., Hoffman, G.L., Kornreich, D.A., Leisman, L., \& Miller, J.R. 2019, AJ, 157, 81 

Phillips, K. W. 2014, Scientific American, 311, 42

Ribaudo, J. et al. 2019, "Primarily Undergraduate Institutions and the
Astronomy Community," State of the Profession White Paper, Astro 2020 Decadal Survey

Rowlett, Roger S., Blockus, L. \& Larson, S. 2012, in \it Characteristics of Excellence in Undergraduate Research, \rm Nancy Hensel (Ed). Washington, D.C.: Council on Undergraduate Research.

Russell, S.H., Hancock, M.P., McCullough, J. 2007, Science. 2007, 316, 548

Sadler, Troy D., Burgin, s., McKinney,, L. 2010, Journal of Research in Science Teaching: The Official Journal of the National Association for Research in Science Teaching, 47.3, 235

Slater, S. J. 2010. \it The Educational Function of an Astronomy REU Program as Described by Participating Women, \rm Ph.D. Thesis, U. Arizona

Troischt, P. W.,  Koopmann, R. A.,  O'Donoghue, A., Odekon, M. c., Haynes, M. P. 2016, Council on Undergraduate Research Quarterly, 36

Wenger, E. 1998 in \it Communities of Practice: Learning, Meaning and Identity, \rm Cambridge: Cambridge University Press, 276

\end{document}